\documentclass{PHYEAUTH}
\usepackage{graphicx}
\usepackage{amsmath}
\usepackage{amssymb}

\def\be{\begin{equation}}
\def\ee{\end{equation}}
\def\bea{\begin{eqnarray}}
\def\eea{\end{eqnarray}}

\begin{document}

\begin{frontmatter}

\title{Quantum dot cascade laser: Arguments in favor. }

\author[address1,address2]{I.A. Dmitriev\thanksref{thank1}}
and
\author[address1]{R.A. Suris}

\address[address1]{Institut f\"ur Nanotechnologie, Forschungszentrum
Karlsruhe, 76021 Karlsruhe, Germany}

\address[address2]{A.F.~Ioffe Physico-Technical Institute,
194021 St.~Petersburg, Russia.}

\thanks[thank1]{
Corresponding author.
E-mail: dmitriev@int.fzk.de}

\begin{abstract}
Quantum cascade lasers are recognized as propitious candidates for future
terahertz optoelectronics. Here we demonstrate several definite advantages of
quantum dot cascade structures over quantum well devices, which suffer
fundamental performance limitations owing to continuous carrier spectrum. The
discrete spectrum of quantum dots opens an opportunity to control the
non-radiative relaxation and optical loss and also provides for more flexibility
in the choice of an optical and electrical design of the laser.
\end{abstract}

\end{frontmatter}

\section{Introduction}\noindent
Owing to active development in the past 10 years, nowadays quantum well cascade
lasers (QW CLs) \cite{faist94,sur71} are the most effective
compact light sources in a broad range of spectrum approaching the terahertz
(THz) frequencies \cite{koehler,williams}, with practical
interest from such diverse sectors as pollution and cruise control, cosmology,
nanotechnology, and medicine. The performance of QW CLs, however, is
fundamentally limited owing to continuous electronic spectrum in QWs, which
leads to fast depletion of the upper laser level by means of LO phonon emission,
as well as high optical loss and strong heating arising from free carrier
absorption.

These fundamental limitations can in principle be avoided if all carriers in a
cascade structure are confined in all three dimensions. That is why an idea to
use quantum dots (QDs) for cascade lasing has raised a lot of interest in recent
years \cite{nato,wingreen,zory,PSS}. In Refs.~\cite{wingreen,zory},
QDs serve mainly for reduction of nonradiative decay rate of the lasing
transition. In other relations the designs of QD CLs proposed there closely
resemble that of QW CL. A rather different design of QD CL, which implements a
dc biased superlattice of coupled QDs, with fully discrete spectrum of carriers,
was proposed in Ref.~\cite{nato}. Albeit substantial technological
difficulties in manufacturing the QD structures, they promise revolutional
improvement of CL characteristics \cite{PSS}, which was also demonstrated
experimentally \cite{scalari}: extremely long nonradiative lifetimes and a
strong reduction of the optical loss were observed in magnetic field for a
specially designed QW CL, which utilized the advantage of nearly discrete
electronic spectrum at well-separated Landau levels localized by disorder. Here
we present more arguments in favor of QD CLs, focusing on the opportunity to
control the nonradiative relaxation in QDs and on possible novel solutions for
the optical and electrical design providing a strong reduction of the optical
loss.

\section{Nonradiative relaxation in quantum dots}

\vspace*{-3mm}\noindent
All existing models \cite{nato,wingreen,zory,PSS} of QD CLs
are built on the assumption of a significant reduction of the nonradiative decay
rate for the lasing transition as compared to QW structures. This assumption is
based on the following: (i) in QWs, intersubband transitions via optical phonon
absorption are very fast and only weakly depend on the energy separation of the
subbands; (ii) only phonons with wavelength $1/q\sim d$ effectively interact
with electrons localized by the potential of a QD of size $d\sim10\,{\rm nm}\gg
a_0$, where  $a_0$ is the lattice constant of the host material, surrounding
QD. It follows that transitions mediated by acoustic phonons are strongly
suppressed for the energy separation $\Delta$ of QD levels exceeding few meV,
while optical phonons with $q a_0\ll 1$ are effectively
dispersionless and thus cannot cause transitions between QD levels.

To make the relaxation in this system possible, we consider two
electron states, $|1\rangle$ and $|2\rangle$, coupled to dispersionless
optical phonons that have a finite lifetime due to anharmonic interaction (AHI)
with all other phonon modes \cite{ahtheories}. In absence of both AHI and
Fr\"ohlich coupling $u_{12}$ of the two states, each QD level forms a polaron
ladder, $|n\nu\rangle$, with the spectrum
$E_{n\nu}=E_{n}+\hbar\Omega(2\nu+1-u_{nn}^2)/2$, where $\Omega$ is the optical
phonon frequency, $n=1,2$, $\nu=0,1,2..$, and $u_{n m}\ll 1$ are Fr\"ohlich
interaction parameters. Account for the coupling $u_{12}$ leads to mixing of the
states of the two ladders. In turn, AHI makes real transitions between the
polaron states possible. To lowest order in $u_{n m}\ll 1$, the rate of
transition from the state $|10\rangle$ reads
\be\label{rate}
W_{|10>\to|2\nu>}=|\langle 1 0|{\bf x}|2\nu\rangle|^2 w(\Delta_\nu).
\ee
Here $w(\Delta_\nu)$ is the probability of
anharmonic decay of a virtual optical phonon accompanied by transfer
of the energy $\hbar\Omega\Delta_\nu=E_1-E_2-\nu\hbar\Omega$ to the phonon bath.
The square of the matrix element of the optical-phonon displacement operator
${\mathbf x}$ is given by \cite{ah}
\be\label{x}
\langle 1 0|{\mathbf
x}|2\nu\rangle^2\!=\!\frac{u_{12}^2(u_{11}-u_{22})^{2\nu}}{2^{\nu-1}
\nu!(\Delta_\nu^2-1)^2}
\left(1+\!\frac{
\nu} {\Delta_\nu}\!\right)^2\!.
\ee
Equations (\ref{rate}), (\ref{x}) demonstrate a strong dependence of
the nonradiative relaxation rate on the energy separation $\Delta=E_1-E_2$ of
the QD levels. In the vicinity of the resonances $|\Delta_\nu|\sim 0, 1$ one
should additionally take into account the avoided crossing of the mixed
polaronic
levels. In particular, the width of the resonance at $\Delta\sim\hbar \Omega$ is
of order $\hbar \Omega |u_{12}|$. Correspondingly, near the resonance with the
optical phonons the relaxation rate $W_{|10>\to|20>}\sim w(1)$ is approximately
the decay rate of the optical phonon in the host crystal, $W_{|10>\to|20>}\sim
w(1)\lesssim10^{12}\,{\rm s}^{-1}$.
By contrast, for the energy separation $\Delta\sim 2.5\hbar \Omega$, which we
use in the following for the laser transition energy, Eqs.~(\ref{rate}),
(\ref{x}) give few orders of magnitude smaller rate, $W(2.5\hbar
\Omega)=W_{|10>\to|22>}+W_{|10>\to|21>}\sim u^6 w(1/2)+u^4 w(3/2)$, where we put
$u_{22}\sim u_{11}\sim u_{12}\equiv u\ll 1$.

\vspace*{-5mm}
\section{Minimal transport model of QD CL}
\begin{figure}[b]
\centerline{
\includegraphics[width=0.6\columnwidth]{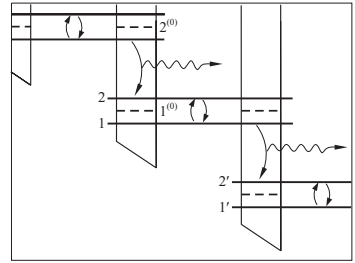}}
\caption{Two-level scheme of QD CL.} \label{fig1}
\end{figure}

\vspace*{-3mm}\noindent
Here we outline the operation principle of a "two-level" scheme of QD CL
\cite{PSS}, as illustrated in Fig.~\ref{fig1}. The active
media of the QD CL utilize chains of coupled QDs \cite{note} subjected to a
strong dc field,
which provides a resonance between the ground state in every QD with the exited
state of the neighboring dot (dashed lines in Fig.~\ref{fig1}). Large tunneling
coupling between the resonant states leads to strong mixing
and splitting. The separation of the splitted levels
is adjusted by applied voltage to the resonance with optical
phonons, $E_{12}\simeq\hbar\Omega$,
providing high transition rate $\gamma\sim10^{12}\,{\rm s}^{-1}$ between them. 
By contrast, scattering time $\tau$ between lasing levels $1$ and $2'$,
separated by $\hbar\omega=2.5\hbar\Omega$, is large, say, $\tau=100\,{\rm ps}$
(see previous section). 
In dynamic equilibrium,  the total occupancy of a dot $n=n_1+n_2$ is
constant along the chain, while the splitted levels at each stage of the
cascade
are nearly in thermal equilibrium: $n_2(1-n_1)=\exp(-\hbar\Omega/T)n_1(1-n_2)$.
As a result, population inversion $\delta=n_1-n_2$ between the lasing levels
occurs:
\be\label{delta}
\delta\!=\!\coth(\hbar\Omega/2T)\!
-\!\sqrt{\!\coth^2(\hbar\Omega/2T)\!-\!n(2\!-\!n)},
\ee
which gives $\delta|_{n=1}\!=\!\coth(\hbar\Omega/4T)\simeq1/3$ at $T=300~{\rm
K}$, where we used $\hbar\Omega=36~{\rm
meV}$ (GaAs).

The current through the chain, $J=e n_1(1-n_2)/\tau$, is
\be\label{cur}
J=(e\delta/\tau)\,[1-\exp(-\hbar\Omega/T)]^{-1}.
\ee
It is important to mention that, unlike QW superlattices \cite{sur74}, in QD CLs
the applied voltage can be adjusted in such a way that the inverse population
occurs on the rising branch of the static current-voltage characteristics, thus
ensuring the electrical stability of the laser operation.
\section{Optical/electrical design}
\begin{figure}[b]
\centerline{
\includegraphics[width=0.8\columnwidth]{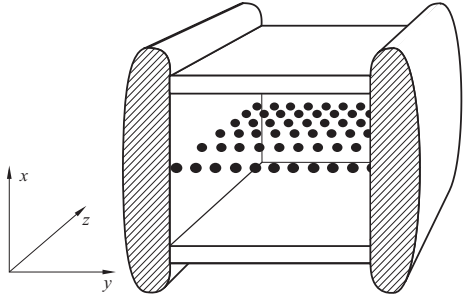}} 
\caption{Design of the ``planar'' QD CL} \label{fig2}
\end{figure}

\vspace*{-3mm}\noindent
In QW CLs, both the electrical component of the laser mode and direction of
injection current should be perpendicular to the QW planes. Thus, only 
TM modes of the cavity can be used. Further, the whole interior of the cavity
should be heavily doped to enable the current densities $\sim {\rm kA}/{\rm
cm}^2$, which results in a high optical loss and strong heating effects. 
By contrast, in QD structures the oscillator strength for the optical transition
is nearly independent on the field orientation, while the free-carrier optical
loss
of QW CLs is absent. 

The sketch of an optical/electrical design that benefits
from the above advantages of QD structures is illustrated in Fig.~\ref{fig2}.
Parallel chains of coupled QDs, oriented along the $y$ direction, form
a lateral QD array in the plane $x=0$, where the electric field 
$E_y(x)$ of the lowest TE mode of the cavity has a maximum.
The optical confinement is provided by the dielectric profile in the $x$
direction.
Side metal contacts at $|y|>w/2$ provide current injection and
operation voltage across the chains, and confine the TE mode in
the $y$ direction. Inside the cavity, the optical loss is negligible, as the
whole structure surrounding QDs is dielectric, so that optical absorption in the
metal near the interfaces becomes the main source of loss. For a good metal with
the surface impedance $\zeta$, the
optical loss, which is associated with the Foucault currents induced by the
magnetic components $H_x$ and $H_z$, is $\beta_{Mt}\simeq{\rm Re}\, \zeta/w$.
Using the Drude approximation for the specific case of palladium
contacts, the laser frequency $\hbar\omega=2.5\hbar\Omega=90~{\rm
meV}$, and the stripe width $w=5\mu{\rm m}$, we get the $T$-independent optical
loss $\beta_{Mt}\sim1~{\rm cm}^{-1}$, which is much
lower than usual internal loss in QW CLs.

We conclude with an estimation of the laser characteristics at the generation
threshold, where the peak modal gain become equal to the total loss:
\be\label{gain}
\Gamma_{\rm th}=16\pi\alpha \,|r|^2\, a\, d\, Q\, \delta_{th}/\lambda=
\beta_{Mt}+\beta_m.
\ee
For the estimate of the threshold inversion $\delta_{th}$ we take the quality
factor of the laser transition $Q=50$, the dipole matrix element of the
transition $r=2~{\rm nm}$, the QD chain period $d=10~{\rm nm}$, the distance
between the chains $a=20~{\rm nm}$, and the mirror loss
$\beta_m=2~{\rm cm}^{-1}$. Using $\alpha=1/137$ and the
operation wavelength $\lambda=14~\mu{\rm m}$, we get $\delta_{\rm th}\sim 0.1$,
which is several times smaller than the achievable value at $T=300K$, see
Eq.~(\ref{delta}). According to Eq.~(\ref{cur}), $\delta_{\rm th}=0.1$
corresponds to the threshold current $J_{\rm th}\sim0.1~{\rm nA}$ per chain,
i.e., to the threshold current density (averaged over the contact area) $j_{\rm
th}=J_{\rm th} 4\sqrt{\epsilon}/a\lambda\sim~0.5{\rm A/cm}^2$. This value is
at least two orders of magnitude smaller than the threshold current density
of QW CLs at $T=300K$.

This work was supported by RFBR Grant No.~05-02-16679, by the Leading
scientific schools support program No.~5730.2006.2, and by the RAS program
``Low-dimensional quantum structures''.


\begin{thebibliography}{8}
\bibitem{ahtheories} Similar mechanisms of the intradot relaxation were studied
previously using different approaches, see \cite{jakak} and references therein.
\bibitem{ah} Generally, a two-level system couples to three
independent Einstein-phonon modes. Here we assume that these
modes coincide, ${\mathbf u}_{12}\|{\mathbf u}_{11}\|{\mathbf u}_{22}$. 
Results for the general case will be presented elsewhere.
\bibitem{note} It is reasonable to expect that inevitable technological
dispersion of QD sizes and distances between QDs, which we do not take into
account here, will be partially compensated by the Coulomb field arising due to
redistribution of electrons in consequence of difference in resistance of
different links in the QD chains.
\end{thebibliography}
\end{document}